\documentclass[10pt,journal,letterpaper,compsoc]{IEEEtran}

\pdfoutput=1

\newcommand{\secref}[1]{\mbox{Section~\ref{#1}}}

\newcommand{\figref}[1]{\mbox{Figure~\ref{#1}}}

\usepackage{latexsym,ifthen,theorem, dsfont, stmaryrd, graphicx, verbatim, listings, amsmath}
\usepackage{url}
\usepackage{mdwlist}
\usepackage{xspace}
\usepackage[center]{subfigure}
\usepackage{epsfig}
\usepackage{color}
\usepackage{verbatim}
\usepackage{placeins}

\begin{document}
\title{A Formal Model for Dynamically Adaptable Services}

\author{Jorge~Fox\thanks{This work was done during the tenure of an ERCIM "Alain Bensoussan" Fellowship Programme}
\IEEEcompsocitemizethanks{\IEEEcompsocthanksitem Jorge Fox is with the Department of Telematics, Norwegian University of Science and Technology, Norway.
\protect\\ E-mail: jfox@item.ntnu.no 
%E-mail: see http://www.michaelshell.org/contact.html
%\IEEEcompsocthanksitem Siobh{\'a}n~Clarke is with Lero -The Irish Software
%Engineering Research Centre at the School of Computer Science
%and Statistics, Trinity College Dublin. \protect\\ E-mail: {firstname.lastname}@cs.tcd.ie 
}% <-this % stops a space
\thanks{}}
\markboth{}%
{Shell \MakeLowercase{\textit{et al.}}: A Formal Model for Dynamically Adaptable Services}

% PRIOR to the title within the \IEEEcompsoctitleabstractindextext IEEEtran
\IEEEcompsoctitleabstractindextext{%
\begin{abstract}
%\boldmath

The growing complexity of software systems as well as changing conditions in their operating environment demand systems that are more flexible, adaptive and dependable. The service-oriented computing paradigm is in widespread use to support such adaptive systems, and, in many domains, adaptations may occur dynamically and in real time. In addition, services from heterogeneous, possibly unknown sources may be used. This motivates a need to ensure the correct behaviour of the adapted systems, and its continuing compliance to time bounds and other QoS properties. The complexity of dynamic adaptation (DA) is significant, but currently not well understood or formally specified. This paper elaborates a well-founded model and theory of DA, introducing formalisms written using COWS. The model is evaluated for reliability and responsiveness  properties with the model checker CMC.

\end{abstract}

}

\maketitle

\IEEEdisplaynotcompsoctitleabstractindextext
\IEEEpeerreviewmaketitle

\section{Introduction}
\label{sec:intro}
Modern software Systems typically operate in dynamic environments and are required to deal with changing operational conditions, while remaining compliant with the contracted quality of service. The execution context of modern distributed systems environments is not static but fluctuates dynamically, and to provide the expected functional service with the desired qualities, systems must be adaptable. Software service adaptation supports modification of existing services or inclusion of new ones, in response to inputs from the operating environment. Inputs or triggers for adaptation include changes in the running environment and availability of new services. When adaptation occurs at runtime, 
quality of service requirements range from timeliness, user perceived performance, service response time, to preservation of data integrity, service performance within given time bounds and the resulting system must comply with the execution time established for the system as a whole. Each one of these requirements has to be addressed accordingly, for instance, preservation of data integrity requires a mechanism to verify that the integrity of the data is kept during and after the adaptation, while timeliness and performance related requirements can be addressed by maintaining the execution times within the desired time bounds.

While the possibility of dynamic service deployment and evolution offers an exciting range of application development opportunities, it also poses a number of complex engineering challenges. These challenges include detecting adaptation triggers, facilitating timely, dynamic service composition, predicting the temporal behaviour of unforeseen service assemblies and preventing adverse feature interactions following dynamic service composition  . Therefore, a dynamically adaptable service has to be able to identify triggers for adaptation, select and conduct an adaptation strategy, and preserve desired quality of service properties, while avoiding undesired behaviours during or after the adaptation.

A promising approach to achieving the required properties in DA services is the use of formal methods and techniques, which have been successfully applied for managing complexity and system development to ensure implementations of high quality \cite{Broy2008}, verification and validation of the behaviour of adaptive programs \cite{1134337}, and architecture engineering processes that validate a program against desired functional properties \cite{DBLP:conf/euromicro/BeekBG09}.
However, most previous efforts to validate dynamically adaptable services against desired functional properties have been either limited to verifying conditions that are defined before runtime execution or impose high verification costs as each adaptation must be individually modelled and verified, making these approaches extremely expensive. Consider that an adaptive program with ${N}$ different behaviours potentially has ${N^2}$ distinct ways of adaptation. The result is that building DA services is either expensive or limited to predefined adaptation scenarios. Even more, these approaches focus their analysis on the resulting program neglecting the analysis of the adaptation mechanism itself, that is the adaptation manager (AM). We consider the adaptation mechanism to be of capital importance in the development of DA services, since it is responsible for monitoring compliance of the adaptation to desired properties and functionalities.

Current approaches to 
DA propose various mechanisms for handling adaptation, such as: Generic Interceptors \cite{10.1109/ICDCS.2004.1281570}, DA with aspect-orientation \cite{582144}, Dynamic Reconfiguration \cite{Pelle03},  Dynamic Linking of Components \cite{EscoffierH07}, and Model-Driven Development of DA Software \cite{1134337}. However, none of these provides a formal framework to verify the adaptation mechanism against commonly-accepted service properties.
Formal languages provide the underpinnings to explain and model software systems in a precise manner, and are fundamental for the level of analysis, validation and proof required for assuring adaptation compliance to specifications. In this paper, we propose an abstract conceptual model of an AM and explore its operation in view of quality of service properties, such as availability and responsiveness.

This article is organised as follows: \secref{sec:overv} provides an overview of current DA approaches. In order to build a formal model of the problem, an underlying language that facilitates its description and depicts the desired functionality properties at the same abstraction level is required, \secref{sec:langs} discusses the steps we followed to select this foundation and the model checking tools supported by the language. \secref{sec:model} 
introduces our formal model and discusses its properties. Availability and responsivenes are identified as desirable attributes of services \cite{Fantechi2008}. In \secref{sec:verif} we explore our model in view of these attributes, and elaborate some lessons learned and topics requiring more research. \secref{sec:relwork} discusses related work. Finally, \secref{sec:conc} presents some conclusions.

%\section{Background}
%\label{sec:bckgd}

\section{An Overview of Dynamic Adaptation Techniques}
\label{sec:overv}

Some existing techniques for DA provide mechanisms to incorporate adaptation elements at design time, such as the work of \cite{Rouvoy2008} or allow for reconfigurations at runtime. These techniques can be classified according to the extent or scope of the adaptations that are made possible, the degree to which the adaptation triggers have to be known in advance, and the tools that the particular framework offers to implement or program a dynamically adaptive software. We also find techniques that allow dynamic linking and unlinking components or services as in \cite{EscoffierH07} and techniques that apply aspect-orientation to achieve DA as \cite{582144}. Another group of techniques offer reconfiguration techniques that enable systems to adjust internal or global parameters to respond to changes in the environment as in \cite{Purtilo1994}, or Generic Interceptors with Adaptive CORBA, which enables message interception to add additional behaviour for adaptation \cite{10.1109/ICDCS.2004.1281570}.

While these techniques offer an interesting range of options to achieve different degrees of DA, questions related to adaptation trigger identification and soundness of a given adaptation model are still open. 

\section{Formal Foundation}
\label{sec:langs}

\subsection{The Service-oriented Language COWS}
\label{sec:cows}
In order to elaborate a model of dynamic adaptation, we  explored a number of languages in \cite{foxclarke2010-iceccs}. Current service-oriented formal languages like: PiDuce \cite{PiDuce09}, {SOCK/JOLIE} \cite{GuidiSOCK06}, {COWS} \cite{LPT07:ICTAC}, KLAIM \cite{Bettini03theklaim}, and SCC \cite{BorealeBCNLLMMRSVZ06} offer each a different range of possibilities to model DA. Every one of these languages has an underlying process algebra and constructs that support defining services and expressing substitution and deactivation processes. The prime mechanism in PiDuce to model service substitution is through virtual machines, while in the case of COWS it is modelled by delimited receiving and killing activities handled with its process calculus, JOLIE and PiDuce offer no deactivation process.
After reviewing these languages, we concluded that the best-fit language for modelling our runtime dynamic adaptation problem is COWS. Given the characteristics of the languages selected, where only COWS provides constructs for timing analysis. Timeliness was not the only deciding criteria, considering only extent adaptation of services merely via channel renaming is sufficient to achieve DA is questionable, as is the case of PiDUce. In this regards the composition mechanisms of SOCK/JOLIE are more adequate, yet again with no possibility to evaluate timeliness. Our choice is clear and well founded. For more on the language COWS the reader may refer to \cite{LPT07:ICTAC}.

\subsection{Model Checking Tools}
\label{sec:modcheck}

Model checking is an automatic technique for verifying finite-state reactive systems.
Specifications are expressed in a propositional temporal logic, and the reactive system
is modeled as a state-transition graph. An efficient search procedure is used to determine
automatically if the specifications are satisfied by the state-transition graph \cite{DBLP:conf/fsttcs/Clarke97}.

Model checking \cite{Clarke2000} as an automatic verification technique covers a wide field of diverse, often ad hoc, and incomplete methods for showing correctness, or, more precisely, for finding bugs. Other verification techniques include theorem proving \cite{Giunchiglia2000} and testing \cite{Pretschner2004}. 
This technique allows software developers to find subtle errors in the design of safety-critical systems that
often elude conventional simulation and testing techniques in a proven cost-effective manner by systematically exploring the state space of concurrent or reactive systems. Furthermore, model checking integrates well with conventional design methods. This is the reason why model checking is being adopted as a standard procedure for the quality assurance of reactive systems. This is done by computing a system's state space from an abstract description specified in a modeling language.
Several properties of a model of a system can then be checked by exploring this state space: deadlocks,
dead code, violations of user-specified assertions, etc. The properties
that state-space exploration techniques can verify has been substantially broadened 
thanks to the development of model-checking methods for various temporal
logics. A number of model checkers are available, such as: SPIN \cite{Holzmann200577}, SMV \cite{SMVManual}, UMC  \cite{DBLP:conf/forte/AbreuMFG09}, and CMC \cite{1060297,1133418}, as well as related analysis tools and languages such as the analyser Alloy \cite{505149}. 

%Conventional model checkers usually assume that
%the design is described at a high level that abstracts
%away many details of the actual implementation.
%Verifying actual code using such a tool requires reconstructing
%this abstract description from the code.
%This process requires a great deal of manual effort, hampering the use of model checking in actual system
%design. Moreover, human errors in the manual
%abstraction process result in missing bugs and cause
%false alarms during verification, further increasing
%the cost and reducing the usefulness of model checking.
%Such errors can be introduced both when constructing
%the model and as a result of "drift" as the
%actual system evolves [5]. For these reasons, it is
%a notable curiosity when software is model checked,
%rather than an everyday occurrence.

%CMC (C Model Checker) to address some of these issues. 
CMC works on unmodified C
or C++ implementations and explores large state
spaces efficiently by storing states. Like traditional
model checkers, CMC achieves the equivalent of executing
astronomical numbers of tests in reasonable
time. 
In this work, we use the model checker CMC \cite{cmc} for the above mentioned reasons and 
its seamless integration with the language COWS.

%NEXT: HERE EXPLAIN THE PROPERTIES WE CAN AND WISH TO CHECK ON OUR MODEL.

%\IEEEpubidadjcol

\section{A Model for Dynamic Adaptation}
\label{sec:model}

A wide range of currently available approaches for services adaptation offer several techniques to achieve DA. However, at the same time most are static, and more importantly,
the flexibility of adaptation or the degree at which adaptations
are achieved, is in most cases limited. Also, in many existing
DA frameworks adaptation is achieved by parametrisation or reconfiguration,
which may render limited solutions with respect to
flexibility and limit further adaptations. 
Another area of opportunity we identify is the need to obtain service adaptation while ensuring 
compliance with predefined properties. This is precisely the motivation behind this work. We introduce in this work a model for DA that has been validated against desirable attributes of services and service-oriented computing applications \cite{DBLP:books/sp/dcsa/Alonso04}. According to this, a service should be: available, reliable, and responsive.

This section presents a two-step process for establishing our DA model. The five steps are listed as follows:
\begin{enumerate}
\item A scenario explaining a dynamic adaptation problem
%\item Identification of key elements for DA and description of their interplay
\item Design of the model in a formal language suitable for verification
\end{enumerate}

\subsection{DA Scenario}
\label{sec:scenario}
\begin{figure}[tb]
\centering
\includegraphics[scale=0.5]{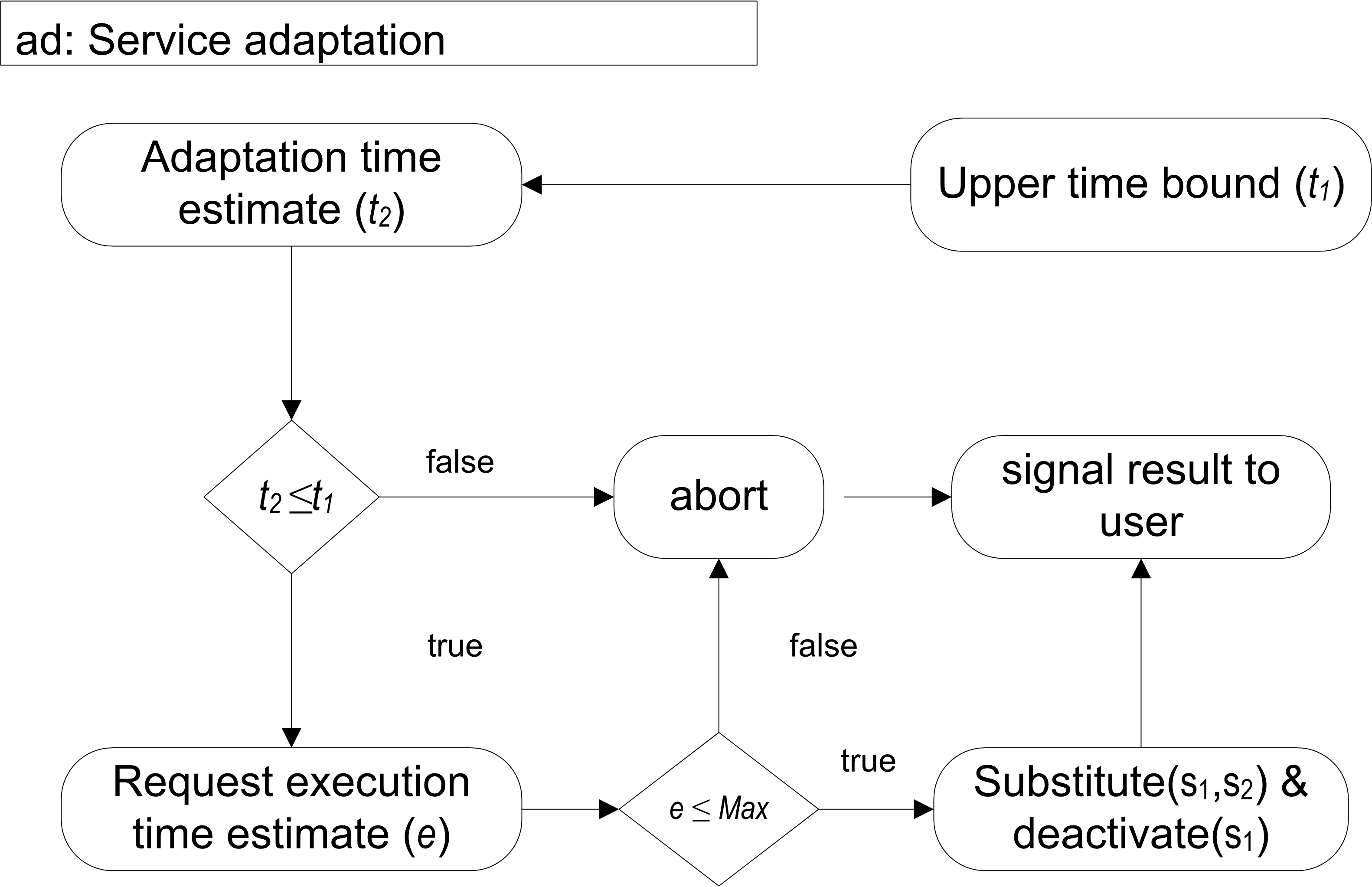}
\caption{Scenario. Service adaptation procedure}
   \label{fig:ServAdap}
\end{figure}

The scenario is based on a tollbooth system, that can be considered a distributed system. The flow of events is given below. In this scenario a car approaches a tollbooth and gets a welcoming signal from it. If the car's payment protocol is not compatible to that of the tollbooth, then it receives a new protocol from the tollbooth. 
Afterwards, the car verifies the protocol and if needed adapts its electronic toll system to comply with the tollbooth's protocol. The scenario is illustrated in \figref{fig:ServAdap}.

The ETS has access to an electronic money pocket that is adjusted to a given tollbooth protocol, when the protocol is available. 
Modifying the system to comply with a new ETS protocol, steps 6-a and 6-b, in the tollbooth scenario, have to be performed at runtime. The adaptation is executed by a service that substitutes the former one. DA is the solution for this real-time adaptation problem.

\begin{figure*}[tb]
\begin{center}
\includegraphics[scale=0.6]{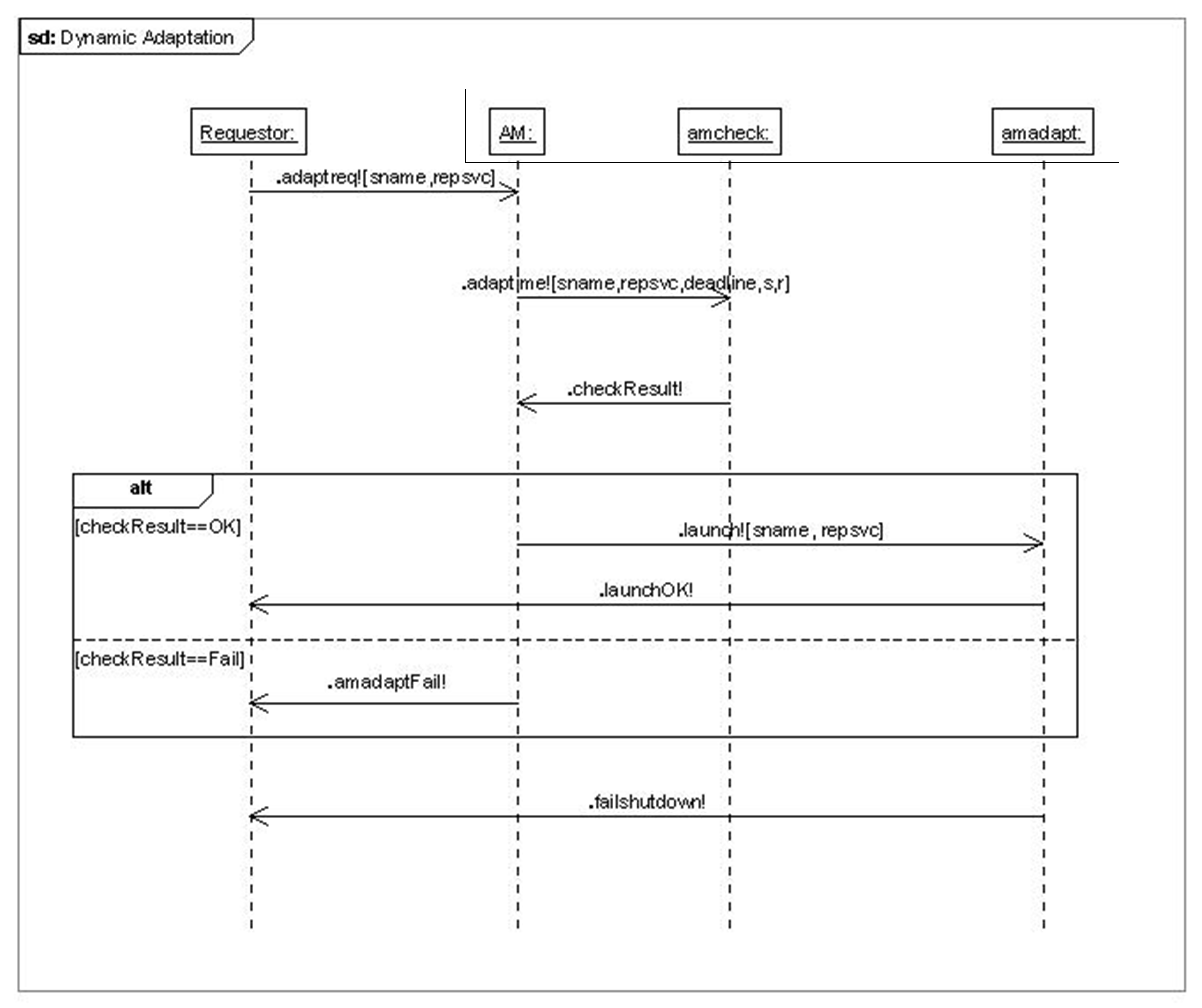} 
\caption{Activity Diagram. Adaptation Process}
\label{fig:DA}
\end{center}
\end{figure*}

\subsection{Model of DA in COWS}
\label{sec:keyE}

The scenario in \figref{fig:DA} indicates the need for a mechanism to identify the adaptation trigger that initiates the adaptation process. We call it ``Requestor'', in our scenario it is represented by a welcoming signal and the request to confirm compatibility with the tollbooth. When the Requestor triggers an adaptation, this request has to be processed by another service, this service is the AM. The trigger for adaptation is processed and a decision on whether to proceed with an adaptation or not is taken also by the AM.
Preservation of timeliness constraints has to be considered and incorporated in the AM, estimating the time to achieve adaptation and integrating this information with predefined upper-bound values. In case the adaptation can be realised within the specified time bounds, the AM may start a reconfiguration process.

%A general overview of this model is shown in \figref{fig:DA}. 
The model illustrates the control flow for the adaptation, starting with the requestor, then a service \texttt{amcheck} is called with the time estimations. The timeliness conditions are examined in \texttt{amcheck} (listing \ref{lst:Atime}), in case the adaptation can be fulfilled within the defined time bound, a second condition is evaluated (listing \ref{lst:Atime2}), which verifies that the adaptation preserves the overall execution time of the service. In case both conditions are uphold \texttt{amadapt} is called and the requestor receives the signal \texttt{signalOK} registering a successful adaptation (listing \ref{lst:AMCOWS}).
Our model focused on two timeliness conditions, adaptation time and execution time, however, other conditions that the system has to preserve can be analysed by the AM in the same manner. For instance, verifying that the service to be reconfigured is in a quiescent state before performing any changes in order to avoid interaction and coordination conflicts.  
%\figref{fig:ServAdap}. 

%\begin{table}
%\begin{table}[tb]

\section{Verifying Quality of Service Properties on the Model}
\label{sec:verif}

Dynamic software architectures allow us to build dynamically adaptable systems by supporting the modification of a system's architecture at runtime. Possible modifications include structural adaptations that change the system configuration graph, for example, the creation of new and the deletion of existing components and connectors, as well as functional adaptations where components are replaced. Further, even more challenging changes are those that modify the behaviour of components and ultimately services.

In order to achieve reliable dynamic adaptable services we need to evaluate service modifications against a number of quality of service properties. Services must comply as a minimum with the following three properties. The first one, Responsiveness, means that the service always guarantees an answer to every received service request, unless the user cancels. The second property, Availability, requires that the service is always capable to accept a request. Finally, the third property, Reliability means that the service request can always succeed.

In order to evaluate a system against these properties, we first have to formulate them in a language capable of being analysed, preferably in an automated manner. 
In the following we introduce our representation of these properties in COWS-CMC. Afterwards, we explore the model at the hand of these properties with the model checker CMC evaluating its compliance to the properties and show a run of the model checker in Appendix \ref{sec:appB}.

In order to explore the first property, Responsiveness, we define a formula with a universal quantifier on the response $amcheck$ following a call to the AM specified as an existential quantifier to the AM, as shown in Listing \ref{lst:conresp}. A run of the model checker can be found in \figref{fig:CMCRelirun}.

%\makebox[10pt] {
\begin{lstlisting}[mathescape,breaklines=true,caption={Responsiveness},
    label={lst:conresp}]
Responsiveness

AG[request(adaptManager)]true
EG[accepting_request(adaptManager)] AF[ response(amcheck)]true

\end{lstlisting}
    
Availability can be easily represented by a universal quantifier on calls to the AM as illustrated in Listing \ref{lst:conavai}. 
%REMAKE

%\mbox{
\begin{lstlisting}[mathescape,breaklines=true,caption={Availability},
    label={lst:conavai}]
Availability
AG[request(adaptManager)]true
\end{lstlisting}
%}

A run of CMC to check Availability can be found in \figref{fig:CMCAvailrun}.

%\mbox {
\begin{lstlisting}[mathescape,breaklines=true,caption={Reliability},
    label={lst:conreli}]
Reliability

EG[request(requestor)] EG[response(launchOK)] EG[response(launchFail)]true
\end{lstlisting}
%}    

An examination of the model with the model-checker CMC confirms that it complies with the third property, Reliability, formulated in Listing \ref{lst:conreli}. See 
 \figref{fig:CMCRelirun}. A run of the model checker can be found in \figref{fig:CMCResponsi}.

%\lstinputlisting[mathescape,breaklines=true,caption={CMC run on Listing \ref{lst:conreli}},
%    label={lst:CMCResp}]{CMCResponsive.txt}

%\section{Discussion}
%\label{sec:disc}

\section{Related Work}
\label{sec:relwork}
We introduce related work in two groups. First, those approaches specifically related to DA, second, formal approaches that are applied to analyse similar families of problems as the ones presented here.

{Dynamic Adaptation}\\
%Mckinley and Geihs (\cite{mckinley-taxonomy,10.1109/MC.2004.48} and \cite{Geihs07}) present an overview of DA and its constituents, however they do not advance a formal model or proposal to explore DA, which is the aims of our work. 
We find an overview of DA and its constituents in the work of Mckinley et al. \cite{10.1109/MC.2004.48,mckinley-taxonomy}, however they do not advance a formal model or proposal to explore DA, which is the aims of our work. 
Similarly to the elements of DA we identified, Segarra and Andr{\'e} describe a similar model to ours with components that can be customized for different applications, a component in their framework can be provided with a controller which performs the adaptation depending on execution conditions \cite{833210}. In our proposal we define one controller, the adaptation manager, that gathers information from supporting services such as timing and execution evaluation in order to perform adaptations.
Work has also been carried out to map BPEL to Process Algebras as Ferrara \cite{1035202}, to Pi-calculus as Abouzaid \cite{1566690}, and to Petri Nets as Ouyang et al. \cite{1274469}.  %WS-BPEL lacks a formal  semantic. %I was wrong
%A comprehensive survey on adaptive systems can be found in \cite{Geihs07,10.1109/MC.2004.48,mckinley-taxonomy}. 
%A comprehensive survey on adaptive systems can be found in \cite{Geihs07,10.1109/MC.2004.48,mckinley-taxonomy}. 

%\subsection*{Formal approaches to DA}
{Formal approaches to DA} \\
The work of Laneve and Zavattaro \cite{Laneve05foundationsof} on web services advances an extension to the $\pi$-calculus with a transaction construct, the calculus web$\pi$. This model supports time and asynchrony. However it remains at a more abstract level and is not applied to dynamic adaptation. Ferrara \cite{1035202} relies on process algebra to design and verify web services, this work also allows to verify temporal logic properties as well as behavioural equivalence between services. Compared to this work, our attempt is more general and is directed at the study of dynamic adaptation. Finally our proposal is aimed at identifying a formal service-oriented language for modelling dynamic adaptation, rather than advancing techniques for formal verification of web services or services as in the work of Ferrara. Mori and Kita \cite{NaokiHaji00} explore the use of genetic algorithms to dynamic environments and offer a survey on problems of adaptation to dynamic environments. 
The work of ter Beek at al. \cite{BeekBucchGnesi07} reviews service composition approaches with respect to a selection of service composition characteristics and helps to underscore the value of formal methods for service analysis at design, specially service composition. The authors present a valuable analysis of formal approaches to service composition and elaborate a useful comparison.
We mentioned the need to provide mechanisms to assure consistency of the system during and after an adaptation, this has been further explored by Amano and Watanabe \cite{NorikiTakuo01}, at this stage, we do not aim at discussing consistency. Nevertheless by relying on a formal language we will be able to support consistency checks. 

\section{Conclusion}
\label{sec:conc}
Real time DA is an area of research that poses new challenges to software development, considering software that may adapt to changing conditions in the operational environment, where new services may be added as they become available, or cope with reconfiguration issues, all this at runtime and under time constraints. DA has been proposed to provide solutions to these challenges. Proposing a methodology for the study of DA is still an open question. Formal methods have been in use for a long time in the computer science community and a number of new approaches and formal languages is available. Modelling DA with a formal language can provide precise answers to most of the existing questions and grant a better understanding of DA. In this work, we explored the use of the formal language COWS to model DA and try our model against three widely accepted service properties: Responsiveness, Availability and Reliability, with the model checker CMC, which integrates seamlessly with COWS.

\ifCLASSOPTIONcompsoc

%\section*{Acknowledgments}
%\else
%  % regular IEEE prefers the singular form
%\fi
%
%
%This work was supported, in part, by Science Foundation Ireland grant
%03/CE2/1303\_1 to Lero - the Irish Software Engineering Research Centre
%(www.lero.ie)

\clearpage
\appendices
\section{Adaptation Manager Model in COWS. Listings}
\label{sec:appA}
To provide some details of the adaptation model,
we introduce here their listings.

\begin{table}[htb]
\begin{lstlisting}[mathescape,caption={\small Adaptation Manager, Requestor and Main Service},
    label={lst:AMCOWS}]

let
adaptManager(service) = 
    * [X][Y][Z][XX][YY]
    service.create?<X,Y,Z,XX>.p.adaptime!<X,Y,Z,XX> 
  | [X][Y][Z][XX]
    ser.checkOK?<X,Y,Z,XX>.q.exectime!<Z,XX>   
  | ser.checkFail?<>. ser.launchFail!<repsvc>
  | ser.checkFail2?<>. ser.launchFail!<repsvc>
requestor() =
    serv.create!<0,4,10,60>  
  | ser.checkOK2?<>.amadapt.launchOK!<> |
    (amadapt.launchOK?<>.s.signalOK!<>
    + ser.launchFail?<repsvc>.s.signalFail!<>
    + ser.launchFailx?<>.s.signalFail!<>)
in
adaptManager(serv) 
| requestor()
| * amcheck()
| amcheck2()
| s.signalFail?<>.nil
| s.signalOK?<>.nil
end
\end{lstlisting}
\end{table}

%\begin{table}
\begin{table}[tb]
\begin{lstlisting}[mathescape,caption={\small Adaptation-time check},
    label={lst:Atime}]
 Amcheck_gt_deadline(X) =
   (ser.checkFail!<>)
 Amcheck_le_deadline(X,Y,Z,XX) =
   (ser.checkOK!<X,Y,Z,XX>)
 Amcheck_gt_deadline2(X) =
    (ser.checkFail2!<>)
   | memory.assert?<X>.nil
 Amcheck_le_deadline2(X) =
   (ser.checkOK2!<>)
amcheck()=
    [X][Y][Z][XX]
    p.adaptime?<X,Y,Z,XX>.
    [i#]         
    (i.selectgreater!<X gt Y> |
      (i.selectgreater?<true>.
       Amcheck_gt_deadline(X) +
       i.selectgreater?<false>.
       Amcheck_le_deadline(X,Y,Z,XX)
      )
    )
    
     
    
\end{lstlisting}
\end{table}

\begin{table}
\begin{lstlisting}[mathescape,caption={\small Execution-time check},
    label={lst:Atime2}]
amcheck2()=
    [X][Y]
    q.exectime?<X,Y>.
    [i#]         
    [K]
    (i.selectgreater!<X gt Y> |
      (i.selectgreater?<true>.
       Amcheck_gt_deadline2(X) +
       i.selectgreater?<false>.
       Amcheck_le_deadline2(X)
      )
     )    
\end{lstlisting}
\end{table}

\newpage

\FloatBarrier

\section{Verification of Properties with CMC}
\label{sec:appB}
The runs of the model checker CMC on the conditions specified in \secref{sec:verif} follows.

\begin{figure}[htb]
\centering

\subfigure[Reliability Check with CMC]{
   \includegraphics[scale =3] {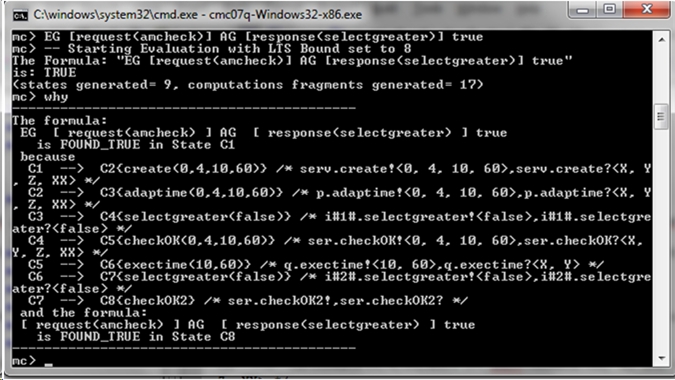}
   \label{fig:CMCRelirun}
 }

 \subfigure[Availability Check with CMC]{
   \includegraphics[scale =3] {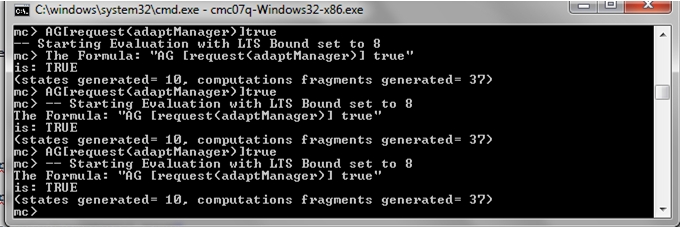}
   \label{fig:CMCAvailrun}
}

 \subfigure[Responsiveness Check with CMC]{
   \includegraphics[scale =2.8] {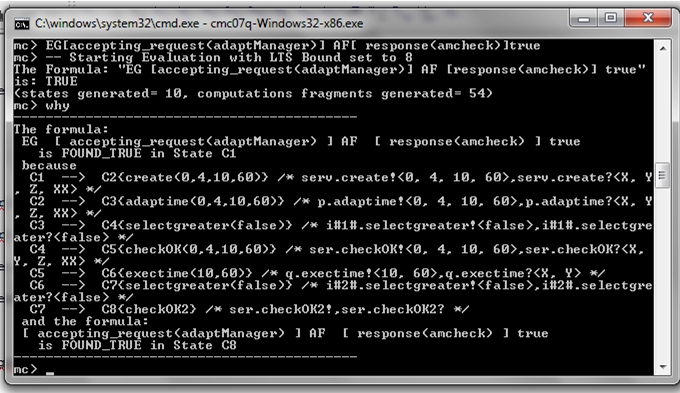}
   \label{fig:CMCResponsi}
 }

\label{myfigure}
\caption{CMC Runs}
\end{figure}

\FloatBarrier

\ifCLASSOPTIONcaptionsoff
  \newpage
\fi

\clearpage
\newpage
\end{document}